\documentclass[floatfix,superscriptaddress,aps,showpacs,twocolumn,twoside,pra,10pt]{revtex4-1}
\usepackage{amsmath} 
\usepackage{amssymb}
\usepackage{graphicx}
\usepackage{xcolor}      
\usepackage{hyperref}   
\hypersetup{
    colorlinks=true,
    linkcolor=blue,
    filecolor=magenta,      
    urlcolor=cyan,
}
\usepackage{natbib}
\setcitestyle{square, comma, numbers, sort&compress}

\usepackage[capitalize]{cleveref}

\def\ket#1{|#1\rangle}
\def\bra#1{\langle#1|}

\def\matr#1#2#3{\left\langle#1\left|#2\right|#3\right\rangle}
\def\abs#1{\left\lvert#1\right\rvert}

\def\={\!=\!}
\def\>{\!>\!}
\def\<{\!<\!}
\def\-{\!-\!}
\def\+{\!+\!}

\def\abs#1{\left|#1\right|}

\newcommand{\nn}{\nonumber\\}

\usepackage{makecell}

\newcommand{\trace}[1]{{\rm Tr}\left\{ #1 \right\}}
\newcommand{\ii}{\mathrm{i}}
\newcommand{\expval}[1]{{\langle #1 \rangle}}

\newcommand{\changes}[1]{#1} 

\begin{document}

\title{Superradiant many-qubit absorption refrigerator}

\author{Michal Kloc}
\email[]{michal.kloc@unibas.ch}

\affiliation{Department of Physics, University of Basel, Klingelbergstrasse 82, CH-4056 Basel, Switzerland}
\author{Kurt Meier}

\affiliation{Department of Physics, University of Basel, Klingelbergstrasse 82, CH-4056 Basel, Switzerland}

\author{Kimon Hadjikyriakos}
\affiliation{Institut für Theoretische Physik, Technische Universit\"at Berlin, Hardenbergstr. 36, 10623 Berlin, Germany}

\author{Gernot Schaller}
\email[]{g.schaller@hzdr.de}
\affiliation{Helmholtz-Zentrum Dresden-Rossendorf, Bautzner
Landstra{\ss}e 400, 01328, Dresden, Germany}

\date{\today}

\begin{abstract}
We show that the lower levels of a large-spin network with a collective anti-ferromagnetic interaction and collective couplings to three reservoirs may function as a quantum absorption refrigerator.
In appropriate regimes, the steady-state cooling current of this refrigerator scales quadratically with the size of the working medium, i.e., the number of spins.
The same scaling is observed for the noise and the entropy production rate.
\end{abstract}

\maketitle

\section{Introduction}

\changes{
With many quantum phenomena observable only at lowest temperatures\changes{~\cite{wiseman2010}}, cooling quantum systems is a task of major importance and requires nanoscale refrigerators that do not disturb the fragile quantum properties~\cite{linden2010a,Lev12,taranto2020,manikandan2020}.
%
%
To maintain the cooling power also in the nano-regime, one may exploit quantum effects in refrigeration~\cite{Bin18,Ali79,Kos13,Uzd15}. 
A considerable amount of literature showing that quantum coherence is able to increase the performance of quantum heat machines has been published~\cite{Scu11,Nie15,Lat20,Sin20,Har20,abah2020}.
It has also been shown that even some fundamental limits of classical thermodynamics can be surpassed~\cite{Scu03,Nie18}.

In finite-stroke quantum heat engines, with the working fluid (WF) being represented by $N$ identical two-level systems that are collectively coupled to heat reservoirs, recent works~\cite{Har15,Nie18b,Klo19,Wat20} have studied specific power boosts proportional to $N^2$.}
The underlying mechanism of this super-extensive scaling of power is related to \textit{superradiance}, originally known from quantum optics~\cite{Dic54},
which can also work in the opposite direction as \textit{superabsorption}~\cite{higgins2014a,yang2021a,zens2021a}.
Finite stroke engines, however, require time-dependent control Hamiltonians, which is experimentally challenging. 
Additionally, studies for simple toy models~\cite{newman2017a} suggest that one should verify that the quantum enhancements gained are not dwarfed by the control work required to couple and decouple the quantum WF from the reservoirs.

An alternative scheme encompasses continuous heat engines~\cite{kosloff2014a}, where the WF is in permanent contact with the reservoirs, such that there are no repeated control costs arising.
For this class of heat engines, stationary heat currents can be directed to accomplish certain tasks, provided that the quantum WFs have been appropriately engineered.
Such a task may be to convert heat to chemical work (electric power).
This application is already possible for two grand-canonical reservoirs and with a single quantum dot (two-level) WF~\cite{esposito2009b}.
Alternatively, for three-terminal systems one may aim to direct the heat currents such that heat is simultaneously entering from the hottest (work) and the coldest reservoir, while the waste heat is dumped into the intermediate temperature (hot) reservoir.
In effect, this provides cooling functionality on the cold reservoir. 
This device is known as quantum absorbtion refrigerator (QAR), and it requires a WF with at least three levels~\cite{linden2010a,Lev12,manikandan2020}.
Although it has been reported that two-level WFs may also function as QARs beyond the regime of weak system-reservoir coupling~\cite{mu2017a} (where the reservoirs in general no longer act additively), it should be noted that such setups -- after appropriate mapping transformations -- can be seen as a multilevel-supersystem that is weakly coupled to residual reservoirs.
Different experimental implementations of QARs have been proposed~\cite{venturelli2013a,hofer2016a,mitchison2019a} and realized~\cite{maslennikov2019a}.
In fact, well-known experimental refrigeration techniques like sideband cooling can also be seen as QAR schemes~\cite{mitchison2016a} 
and furthermore, a QAR bears strong similarities to a reversely driven Scovil-Schulz-DuBois heat engine~\cite{scovil1959a}.
\changes{A pedagogical introduction to QARs and description of their physical implementations  can be found in Chapter 6 of~\cite{Bin18}.}

In this paper we demonstrate that a collective quantum advantage -- known from the super-extensive scaling of superradiant decay~\cite{Dic54} or steady-state superradiance~\cite{meiser2010a,vogl2011a} -- can be achieved in the stationary cooling current of a QAR.
Our WF requires an all-to-all interaction between the qubits and collective couplings to all reservoirs and in the perfect limit allows to cool the coldest reservoir down to a temperature that is at lowest $1/3$ the temperature of the intermediate reservoir. 
In contrast to previous approaches~\cite{correa2014a,Cor13,Kil18,manzano2019a,holubec2019,naseem2020a}, our setting is -- at least in principle -- arbitrarily scalable and allows for a simple analytic limit.
To demonstrate the mere attainability and scaling of cooling, we compute the stationary current using methods of full counting statistics~\cite{segal2018a,friedman2019a}.
Additionally, to discuss the reliability of the collective QAR, we also address the noise~\cite{novotny2004a,flindt2004a} by investigating the thermodynamic uncertainty relation~\cite{barato2015a,gingrich2016a} for our system.

This paper is organized as follows:
We start below in Sec.~\ref{SEC:Model} with an introduction to the model and afterwards discuss the basic properties of the Pauli-type master equation in Sec.~\ref{SEC:master_equation}.
We present our results in Sec.~\ref{SEC:Num} before concluding in Sec.~\ref{SEC:summary}.
Details like specific limits of our model for single, two, and three reservoirs or simplified reduced versions in appropriate parameter regimes as well as technical methods are highlighted in the appendices.
Throughout the paper we use the convention $\hbar=1$ and $k_B=1$ for the Planck and Boltzmann constants, respectively. 
Additionally, we note that all plotted quantities are dimensionless.


\section{Model}\label{SEC:Model}

We consider the WF formed by a $N$ two-level systems with an all-to-all anti-ferromagnetic interaction 
\begin{align}
    H_S = \Omega J_z^2\,,
    \label{EQ:Hs}
\end{align}
where the collective spin operator $J_\nu = \frac{1}{2} \sum_{i=1}^N \sigma^i_\nu$ (we will discuss the case of odd $N$) can be expressed with the usual Pauli matrices and where the factor $\Omega$ defines the energy scale of the model.
A corresponding effective Hamiltonian can be implemented by laser-driven ions in a trap~\cite{moelmer1999a} and has experimentally been used for entanglement generation in small systems~\cite{sackett2000a}.

The system is coupled via the interaction Hamiltonians 
\begin{align}
    H_{I,\nu} &= A_\nu \otimes  B_\nu\,,\nn
    A_h &= \frac{J_x^2}{N}\,, \
    A_c = J_x\,, \
    A_w = J_x\,.
    \label{EQ:CouplingOperators}
\end{align}
to a hot ($\nu=h$), cold ($\nu=c$) and work ($\nu=w$) reservoir that are modeled as independent harmonic oscillator baths $H_{B,\nu}=\sum_k \omega_{k,\nu} b_{k\nu}^\dagger b_{k\nu}$ with bosonic annihilation operators $b_{k\nu}$ each.
We note that experimental control of such collective $J_x$-couplings has been demonstrated~\cite{leroux2010a,schleier_smith2010a,dalla_torre2013a}. 
The squared coupling operator can be understood  via $J_x^2 = \sum_{ij} \sigma^x_i \sigma^x_j/4$ as a two-particle flip process~\cite{moelmer1999a,garziano2016a,munoz2020a,ren2020a} and appears more challenging to implement.
In our model, it is required to  drive different transitions via this reservoir.
Its additional scaling with $1/N$ is not essential but included in order to maintain a balanced dissipation strength among all three reservoirs in the thermodynamic limit $N\to\infty$.

Additionally, we remark that for our purposes the work reservoir could be represented by a laser but in order to keep the discussion uniform we will consider all reservoirs to be thermal and described by inverse temperatures obeying the hierarchy $\beta_c\ge \beta_h \ge \beta_w$.

Due to the collective couplings, the global Hamiltonian of the universe $H=H_S+\sum_\nu H_{I,\nu}+\sum_\nu H_{B,\nu}$ conserves the 
total angular momentum operator $J^2=J_x^2+J_y^2+J_z^2$ and the $x$-parity operator~\footnote{The hermitian and unitary parity operator obeys $\Pi^\dagger \Pi = \Pi^2=\mathbf{1}$ and furthermore transforms $\Pi J_x \Pi = +J_x$,  $\Pi J_y \Pi = -J_y$ and $\Pi J_z \Pi = -J_z$, which can be seen by considering the individual spin transformations separately. From this it follows that the parity operator $\Pi$ commutes with $H_S$, $H_{I,\nu}$, and $J^2$.}
$\Pi = e^{\ii \pi (J_x-N/2)}$ of the system, i.e., one finds  $[H, J^2]=0$, $[H, \Pi]=0$, and also $[J^2,\Pi]=0$.
Thus, we can constrain the considerations to the subspace of maximum angular momentum and positive parity, irrespective of the particular perturbative scheme that we use below.
In this sector, which has for odd $N$ in total only $(N+1)/2$ states $H_S \ket{v_m^+} = E_m \ket{v_m^+}$ we have non-degenerate system energy eigenstates with
\begin{align}
E_m &= \Omega m^2 \qquad:\qquad m\in\{1/2,3/2,\ldots,N/2\}\,,\nn
\ket{v_m^+} &= \frac{1}{\sqrt{2}} \left[\ket{j,m}+\ket{j,-m}\right]\,.
\label{EQ:vs}
\end{align}
\changes{Here, the $\ket{v_m^+}$ form the {\em proper collective basis} for our problem, that can be expressed by the usual collective} Dicke states obeying $J_z \ket{j,m}=m \ket{j,m}$ and $J^2 \ket{j,m}=j(j+1)\ket{j,m}$.
\changes{From the local basis (also known as computational basis in quantum information context~\cite{nielsen2000}), the maximum angular momentum sector with $j=N/2$ of the collective Dicke basis can be constructed by acting with raising operators $J_+=J_x+\ii J_y$ on the local basis state $\ket{1\ldots 1}=\ket{j=N/2,m=-N/2}$ multiple times.}

\begin{figure}[ht]
    \includegraphics[width=8cm]{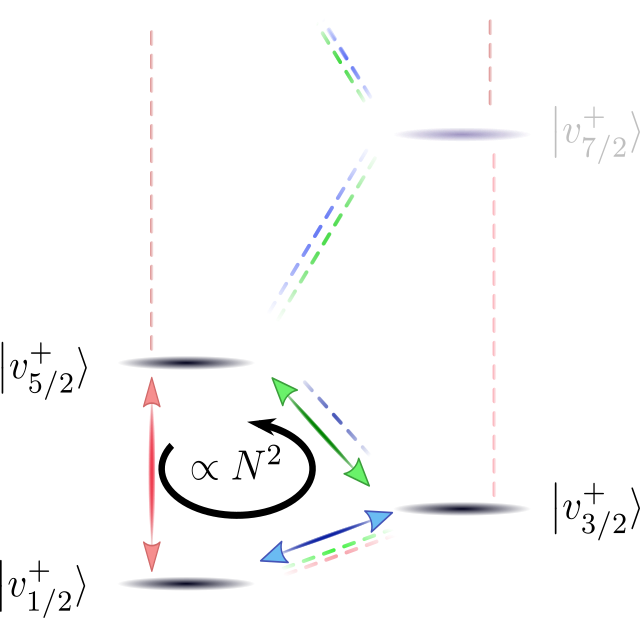}
    \caption{
    The Pauli-type transition rates (arrow-headed solid and dashed lines) generated by the reservoirs~\eqref{EQ:CouplingOperators} connect the energy eigenstates (horizontal levels) of the maximum angular momentum and even parity sector of our system~\eqref{EQ:Hs} ($N$ is assumed odd). 
    In suitable parameter regimes, the lowest three levels of this sector can be driven exclusively by the reservoirs (arrow-headed solid lines). For thermal reservoirs, all transitions prefer the downward direction. However, if this imbalance is for the hot reservoir significantly stronger than for the cold and work reservoirs, the system will on average traverse the cycle counter-clockwise, absorbing energy from the cold reservoir (cooling functionality).
    Since near the system's ground state all rates are boosted superradiantly, the net cooling current is then boosted as well.
    }
    \label{fig:absorbtion_refrigerator}
\end{figure}

In this article, we will consider the situation of odd $N$, where $m$ assumes half-integer values.
Treating only the even parity sector of maximum angular momentum with $j=N/2$ is significantly less demanding than the full set of $2^N$ states that would need to be considered -- together with their coherent superpositions -- if we worked in the local basis instead.
Furthermore, the even and odd parity sectors are for odd $N$ fully symmetric to each other
\footnote{\label{FT:even_N}
If we would choose $N$ even, then for $m\in \{1,\ldots,N/2\}$ we can define vectors $\ket{v_m^+}=\frac{1}{\sqrt{2}} \left[\ket{j,m}+\ket{j,-m}\right]$ in the same way as as for odd $N$ but denote $\ket{v_0^+}\equiv \ket{j,0}$.
The even parity sector would then have dimension $N/2+1$ while the odd parity sector, spanned by vectors $\ket{v_m^-} = \frac{1}{\sqrt{2}} \left[\ket{j,m}-\ket{j,-m}\right]$ with $m\in \{1,\ldots,N/2\}$, would have dimension $N/2$, and the lowest transitions in both sectors would be different.
For simplicity we only consider therefore odd $N$ with even parity in the main text of this article.
}, 
\changes{which we also discuss in App.~\ref{APP:fragility}.}
Even more, the fact that the spectrum in this sector is non-degenerate, allows for a simple rate equation treatment between the relevant energy eigenstates as shown in Fig.~\ref{fig:absorbtion_refrigerator}.
Below, we outline the actual calculation of the transition rates.
\changes{More on the preparation of the specific initial states of interest and their fragility is also discussed in App.~\ref{APP:fragility}.}


\section{Master equation}\label{SEC:master_equation}

\subsection{Reservoir properties}

If the coupling operators are linear $B_\nu = \sum_k h_{k\nu} b_{k\nu} + {\rm h.c.}$ with spontaneous emission amplitudes $h_{k\nu}$, we can write the Fourier transforms of the reservoir correlation functions as
\begin{align}
    \gamma_{\nu}(\omega) &= \frac{1}{2\pi} \int d\tau \trace{e^{+\ii H_S \tau} B_\nu e^{-\ii H_S \tau} B_\nu \bar\rho_B^\nu} e^{+\ii\omega\tau}\nn
    &=     
    \Gamma_{\nu}(\omega) [1+n_{\nu}(\omega)]
    \label{EQ:small_gamma}
\end{align}
with the spectral density $\Gamma_\nu(\omega) = 2\pi \sum_k \abs{h_{k\nu}}^2\delta(\omega-\omega_{k\nu})$ analytically continued to the complete real axis as an odd function $\Gamma(\omega)= -\Gamma_\nu(-\omega)$ and with the Bose-Einstein distribution 
$n_\nu(\omega)=[e^{\beta_\nu\omega}-1]^{-1}$.
We assume that the spectral densities of all three reservoirs can be parametrized as
\begin{align}\label{EQ:specdens}
    \Gamma_\nu(\omega) = \frac{4 \bar{\Gamma}_\nu \varepsilon_\nu \delta_\nu^2 \omega}{\left[(\omega-\varepsilon_\nu)^2+\delta_\nu^2\right]\left[(\omega+\varepsilon_\nu)^2+\delta_\nu^2\right]}\,,
\end{align}
which for $\delta_\nu \ll \varepsilon_\nu$ roughly has a maximum at $\omega=+\varepsilon_\nu$ of height $\bar{\Gamma}_\nu$.
Below, we will be specifically interested in highly peaked spectral densities \changes{which will allow us to dominantly drive only selected transitions in the WF~\cite{friedman2019a,naseem2020a} (see the first paragraph of  Sec.~\ref{SEC:Num})}.
We just note here that such peaked spectral densities can for each reservoir be effectively implemented by coupling the large spin via $\lambda_\nu (b_\nu+b_\nu^\dagger)$ to a single bosonic mode with Hamiltonian $\Omega_\nu b_\nu^\dagger b_\nu$, which is then coupled to its respective residual reservoir, see App.~\ref{APP:specdens_mapping}.

\subsection{Populations in the energy eigenbasis} 

Under the standard Born-Markov and secular approximations~\cite{breuer2002} we can now derive a Lindblad master equation that respects conservation of the collective angular momentum and parity.
Here, one should note that the exact degeneracies of~\eqref{EQ:Hs} in the local ($\sigma_z^i$) basis would induce a coupling of coherences and populations in this basis.
It is then much more convenient to choose \changes{the proper collective basis} $\{\ket{v_{j,m}^\pm}\}$ as representation basis, which is the common eigenbasis of $J_z^2$, $J^2$ and $\Pi$.
Since in the subspace of maximum angular momentum and positive parity the system Hamiltonian~\eqref{EQ:Hs} has no degeneracies, we get a simple Pauli-type rate equation~\cite{mandel1995,breuer2002} for the eigenstates of the system Hamiltonian $H_S \ket{v_a^+} = E_a \ket{v_a^+}$, where $a \in\{1/2,3/2,\ldots,N/2\}$ assumes positive half-integer values only \footnote{From now on we will use indices $a$ and $b$ for the basis~(\ref{EQ:vs}) to avoid confusion with the $J_z$ eigenvectors $\ket{j,m}$.}.
The evolution of these eigenstates is closed in this subspace (see e.g. Sec.~3.3.2 of Ref.~\cite{breuer2002}), meaning that populations only couple to populations according to 
\begin{align}
    \dot{\rho}_{aa} = \sum_b R_{ab} \rho_{bb} - \sum_b R_{ba} \rho_{aa}\,,
    \label{EQ:RateEquation}
\end{align}
where $\rho_{aa}=\bra{v_a^+}\rho\ket{v_a^+}$ are the diagonal elements of the system density matrix in the relevant sector and $R_{ab}$ are transition rates from $\ket{v_b^+}$ to $\ket{v_a^+}$.
The coherences in this basis (if initially present at all), in the long-term limit considered, will just die out. 
Therefore, in the following, we will simply write the populations of the system density matrix as a vector and thereby write the above rate equation as $\dot\rho = R \rho$.
It is important to note that for the assumed weak system-reservoir coupling the matrix elements of the rate matrix $R$ enter additively:
Its off-diagonal  $a\neq b$ elements are given by the rates $R_{ab}=\sum_\nu R_{ab}^\nu$ with 
\begin{align}
    R_{ab}^\nu = \gamma_{\nu}(E_b-E_a) \abs{\bra{v_a^+} A_\nu \ket{v_b^+}}^2\,.
    \label{EQ:SuperradRate}
\end{align}
To evaluate the relevant matrix elements of the coupling operators~\eqref{EQ:CouplingOperators} it is helpful to use that for $a\neq b$ and odd $N$ we can evaluate the matrix elements as
\begin{align}
 \matr{v_a^+}{J_x}{v_b^+} & = \frac{1}{2} \left[
  \alpha_b^+ \delta_{a,b+1} + \alpha_b^- \delta_{a,b-1}
  \right]\,,\nn
  \matr{v_a^+}{J_x^2}{v_b^+} & = \frac{1}{4} \big[
  \alpha_{b+1}^+ \alpha_b^+ \delta_{a,b+2} + \alpha_{b-1}^- \alpha_b^- \delta_{a,\abs{b-2}}
  \big]\,,
\end{align}
where the first matrix element enables the blue and green transitions and the second one the red transitions in Fig.~\ref{fig:absorbtion_refrigerator}, including the additional $\ket{v_{1/2}^+}\leftrightarrow \ket{v_{3/2}^+}$ transitions.
Here, the factors
\begin{align}
  \changes{\alpha_a^{\pm}=\sqrt{j(j+1)-a(a \pm 1)}}\,,
  \label{EQ:Clebsch}
\end{align}
\changes{with $j=N/2$ for the maximum angular momentum subspace}
are the well-known Clebsch-Gordan coefficients from action of the ladder operators on the $J_z$ eigenbasis $J_\pm \ket{a} = \alpha_a^\pm \ket{a}$.
Therefore, the coefficients in the rate equation all scale as $N^2$ for $a \ll N/2$ which implies that the rates close to the ground state of our model~\eqref{EQ:Hs} will be enhanced by this factor [remember that the $J_x^2$ coupling is reduced by factor $N$, cf.~Eq.~\eqref{EQ:CouplingOperators}].
\changes{In App.~\ref{APP:SuperradThermalization} we demonstrate such a scaling in the transient superradiant relaxation due to a a single low-temperature reservoir.
Similarly, the supertransmittance observed for two low-temperature reservoirs is discussed in App.~\ref{APP:supertransmittance}. }

If the coupling of the system to the reservoirs is engineered as described in Sec.~\ref{SEC:Model} we can tune the parameters in such a way that the dynamics take place dominantly among the three lowest lying levels.
This allows us to construct an absorption refrigerator by driving transitions between these levels such that it pumps energy from the cold reservoir to the hot one, see Fig.~\ref{fig:absorbtion_refrigerator}.
Because the respective rates are enhanced by the Clebsch-Gordan coefficients, one can expect a scaling $\bar{I}_E^c \propto N^2$ of the stationary energy current from the cold reservoir as well.

Thus, for a proper initialization in the even parity and maximum angular momentum sector, the time-dependent density matrix is given by the statistical mixture $\rho(t) = \sum_{a=1/2}^{N/2} \rho_{aa}(t) \ket{v_a^+}\bra{v_a^+}$.
Despite the simple diagonal form in the \changes{proper collective} basis, our model features highly coherent states in the local $\sigma^z_i$-basis of the individual two-level systems \changes{(also termed computational basis)}.
To see this, consider e.g. the ground state in the maximum angular momentum and even parity sector. 
A state like $\ket{v_{1/2}^+}\bra{v_{1/2}^+}$ has no coherences in \changes{the proper collective} basis but considering e.g. $N=3$ we see that in the local basis it is highly coherent $\ket{v_{1/2}^+}=\frac{1}{\sqrt{6}}\left[\ket{001}+\ket{010}+\ket{100}+\ket{110}+\ket{101}+\ket{011}\right]$.
These local coherences are the essential ingredient to the quantum speedup we investigate below.

\subsection{Current and Noise}

To obtain the long-term statistics of transferred energy quanta between the system and its reservoirs (in particular, the cold one), we employ the counting field formalism.
We generalize the rate matrix $R(0)=R\to R(\chi)$ by introducing an energy counting field $\chi$ in its off-diagonal elements.
This can of course be done for all reservoirs but we exemplify it only for energy exchanges with the cold one
\begin{align}
    R_{ab}(\chi) &= R_{ab}^h+R_{ab}^w+R_{ab}^c e^{+\ii\chi(E_a-E_b)} \qquad:\qquad a\neq b\nn
    R_{aa}(\chi) &= - \sum_{b\neq a} R_{ba}(0)\,.
\end{align}
For $\chi=0$, we reproduce the previous case where the matrix elements in all columns of the rate matrix have to add up to zero (trace conservation).
The sign of the counting field has been fixed to follow the convention that currents count positive when they increase the energy of the system (and negative otherwise).

The stationary energy current entering the system from the cold reservoir $\changes{\bar{I}_E^c = -\lim\limits_{t\to\infty} \frac{d}{dt} \expval{H_{B,c}}}$ can then be obtained by computing the derivative of the rate matrix with respect to the counting field
\begin{align}\label{EQ:energy_current}
    \bar{I}_E^c = -\ii \trace{R'(0) \bar\rho}\,,
\end{align}
where the stationary state $\bar\rho$ is obtained by solving the matrix equation 
\begin{align}\label{EQ:steady_state}
    \left(\begin{array}{ccc}
    &|&\\
    -& R(0) &-\\
    &|&\\
    1 & \ldots & 1
    \end{array}
    \right)
    \left(\begin{array}{c}
    |\\
    \bar\rho\\
    |
    \end{array}\right) 
    = 
    \left(\begin{array}{c}
    0\\
    \vdots\\
    0\\
    1\end{array}\right)\,. 
\end{align}
Note that in Eq.~\eqref{EQ:steady_state} the last added row enforces the requirement $\trace{\bar{\rho}}=1$.
Alternatively, it may be obtained without explicit calculation of the stationary state by the cofactor matrix of the rate matrix~\cite{friedman2019a}.

The long-term fluctuations (noise) of this energy current
$\changes{\bar{S}_E^c = \lim\limits_{t\to\infty}\frac{d}{dt} \left[\expval{H_{B,c}^2}-\expval{H_{B,c}}^2\right]}$
can be obtained via~\cite{benito2016b,restrepo2019a}
\begin{align}\label{EQ:energy_noise}
    \bar{S}_E^c = -\trace{R''(0)\bar\rho}-2\trace{R'(0)\bar\sigma}\,,
\end{align}
where the auxiliary quantity $\bar\sigma$ can be determined by solving the matrix equation
\begin{align}\label{EQ:steady_sigma}
    \left(\begin{array}{ccc}
    &|&\\
    -& R(0) &-\\
    &|&\\
    1 & \ldots & 1
    \end{array}
    \right)
    \left(\begin{array}{c}
    |\\
    \bar\sigma\\
    |
    \end{array}\right) 
    \= 
    \left(\begin{array}{c}
    |\\
    \left[\trace{R'(0)\bar\rho}\cdot\mathbf{1}\-R'(0)\right]\bar\rho\\
    |\\
    0\end{array}\right).
\end{align}
In App.~\ref{APP:current_noise} we detail the derivation of these formulas.

\subsection{Thermodynamic consistency}

It can be shown that the approach is thermodynamically consistent.
For example, the first law of thermodynamics at steady state is reflected in the fact that
\begin{align}\label{EQ:first_law}
    \bar{I}_E^c+\bar{I}_E^h+\bar{I}_E^w=0\,.
\end{align}
The existence of a second law can be deduced from the fact that the rates obey local detailed balance
\begin{align}
    \frac{R_{ab}^\nu}{R_{ba}^\nu} = e^{-\beta_\nu(E_a-E_b)}\,,
\end{align}
such that the standard thermodynamic framework of Pauli-type master equations~\cite{schnakenberg1976a} applies.
From this one can derive many implications, for 
example that the individual rate matrices locally thermalize the system
\begin{align}
    \sum_b R_{ab}^\nu e^{-\beta_\nu \Omega b^2} = 0\,,
\end{align}
and that the stationary irreversible entropy production rate is positive
\begin{align}\label{EQ:second_law}
    \bar{\sigma}_\ii = -\beta_c \bar{I}_E^c-\beta_h \bar{I}_E^h-\beta_w \bar{I}_E^w \ge 0\,,
\end{align}
which -- in the regime where the Pauli-type rate equation is valid -- constitutes the second law of thermodynamics and bounds the coefficient of performance of our engine by the corresponding Carnot bound, see the discussion in App.~\ref{APP:carnotbound}.
Related, one finds that a Crooks heat exchange fluctuation theorem~\cite{crooks1999a} is obeyed and furthermore, that the standard thermodynamic uncertainty relation (TUR)~\cite{barato2015a,gingrich2016a} holds, which e.g. for the current and its fluctuations from the cold reservoir reads
\begin{align}\label{EQ:turbound}
    \frac{\bar{S}_E^c \bar{\sigma_\ii}}{\left(\bar{I}_E^c\right)^2} \ge 2\,.
\end{align}

The l.h.s. of this equation can be seen as a measure for the reliability of the engine: Its fluctuations cannot be arbitrarily small but follow a universal bound.
Additionally, the above equation can -- analogous to Ref.~\cite{pietzonka2018a} -- be used to bound the coefficient of performance (see App.~\ref{APP:carnotbound})
in the regime where  $\bar{I}_E^c>0$, $\bar{I}_E^w>0$ and $\beta_c\ge\beta_h\ge\beta_w$) as $\kappa \le \bar{\kappa}$ where
\begin{align}
    \bar{\kappa} = \kappa_{\rm Ca} \frac{1}{1+2\frac{\bar{I}_E^c}{\bar{S}_E^c(\beta_c-\beta_h)}}\,,
    \label{EQ:kbar}
\end{align}
which can be a significantly tighter restriction than the standard Carnot bound $\kappa_{\rm Ca}=\frac{\beta_h}{\beta_c-\beta_h}$.


\section{Numerical results}\label{SEC:Num}

The coupling operators~\eqref{EQ:CouplingOperators} and the spectral densities~\eqref{EQ:specdens} necessarily allow for energy transitions within the system also beyond the three lowest-lying levels.
In the ideal limit of highly peaked ($\delta_\nu \to 0$) and resonantly tuned ($\varepsilon_c=2\Omega$, $\varepsilon_h=6\Omega$, $\varepsilon_w=4\Omega$) spectral densities~\eqref{EQ:specdens} for all three reservoirs, the dynamics our model would (if e.g. initialized in the ground state) be fully restricted to the three lowest energy levels (see Fig.~\ref{fig:absorbtion_refrigerator}) while the remaining transition rates would still be boosted superradiantly~\eqref{EQ:SuperradRate}.
This case, which we refer to as the \textit{reduced model}, also allows for analytic treatment as the dynamics is described by a $3 \times 3$ matrix, see App~\ref{APP:reduced_model}.

In this section we present a numerical study of the performance of the model and identify useful working regimes with respect to its parameters.
The analysis is restricted to the maximal $j$ and positive parity subspaces of~\eqref{EQ:Hs} for odd $N$.

\begin{figure}[t]
    \includegraphics[width=1\linewidth,clip=true]{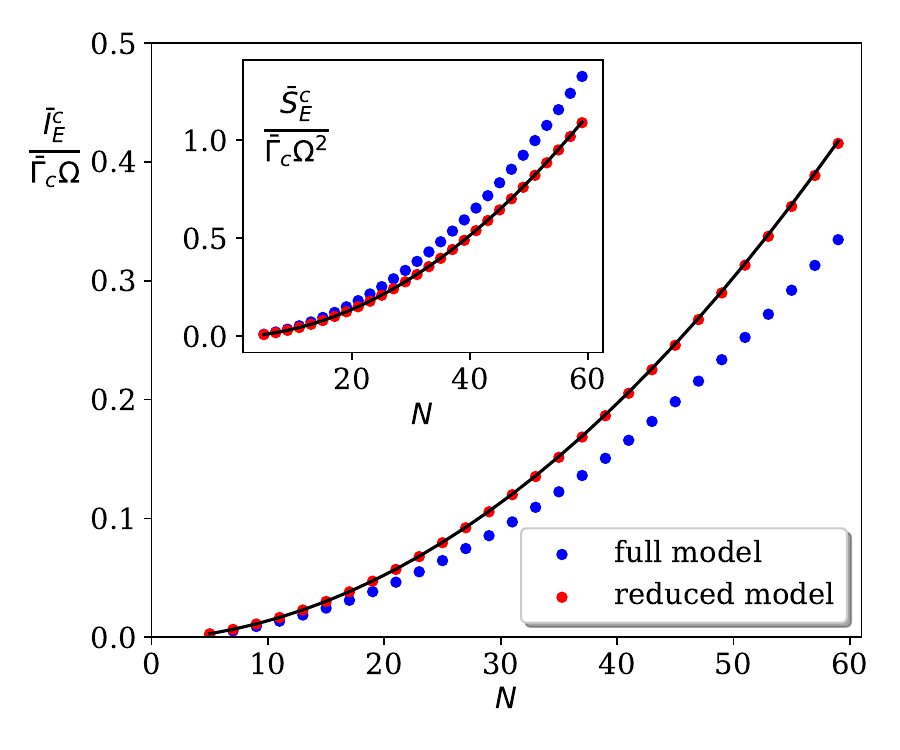}
    \caption{Stationary current $\bar{I}_E^c$ and long-term fluctuations $\bar{S}_E^c$ (inset)  as a function of the (odd) WF size $N$ (blue), see Eq.~\eqref{EQ:Hs},  together with the reduced three-level model (red).
    The solid lines correspond to the analytic formulas~\eqref{EQ:AnalIc} and~\eqref{EQ:AnalSc} for an infinite temperature work reservoir $\beta_w \to 0$.
    Parameters used are 
    $\varepsilon_c=2\Omega$ $\varepsilon_h=6\Omega$, $\varepsilon_w=4\Omega$,
    $\delta_c=\delta_h=0.1\Omega$, $\delta_w=10^{-3} \Omega$, $\beta_c=2 \Omega^{-1}$, $\beta_h=1 \Omega^{-1}$, $\beta_w=10^{-3}\Omega^{-1}$, $\bar{\Gamma}_h=\bar{\Gamma}_w=\bar{\Gamma}_c=1\Omega$.}
    \label{fig:I}
\end{figure}

In Fig.~\ref{fig:I} we show numerical results for the  scaling of the stationary current $\bar{I}_E^c$ and noise  $\bar{S}_E^c$ with $N$ (inset).
The results for the reduced model (see App.~\ref{APP:reduced_model}) and $\beta_w=10^{-3}\Omega^{-1}$ agree almost perfectly with the analytic formulas (solid black lines) for $\beta_w\to 0$ from Eqns.~\eqref{EQ:AnalIc} and~\eqref{EQ:AnalSc}.
All curves show the $N^2$ dependence.
The performance of the full model is a bit worse due to population leakage to the higher energy levels of the WF, which do not productively contribute to the cooling current.
The super-extensive scaling of the cooling current, however, is still present as a major fraction of the populations still inhabitates the lowest three eigenstates.

When measuring the heat current absorbed from the cold reservoir $\bar{I}_E^c$ for a fixed time $\delta t$ (that is large enough to be in the stationary regime), the  transferred energy $\delta E_c$ fluctuates due to the noise $\bar{S}_E^c$. 
The ratio of the width of the energy fluctuations compared to the energy transferred is dimensionless and scales with $N$ as
\begin{align}
   \frac{\sqrt{\bar{S}_E^c \delta t}}{\bar{I}_E^c \delta t} \propto \frac{1}{N}\,.
   \label{EQ:Nsuppression}
\end{align}
Note that in the case of $N$ independent three-level systems forming the WF the quantity~\eqref{EQ:Nsuppression} would follow the $1/\sqrt{N}$ dependence.
Thus, the speedup by the collective interaction also allows to suppress the noise faster, as can be understood by simple intuition:
As the dynamics is boosted, within a fixed time window $\delta t$ the WF runs through the cycle (as marked in Fig.~\ref{fig:absorbtion_refrigerator}) more times for large $N$ so the noise gets averaged out better.

\begin{figure}[ht!]
    \includegraphics[width=1\linewidth,clip=true]{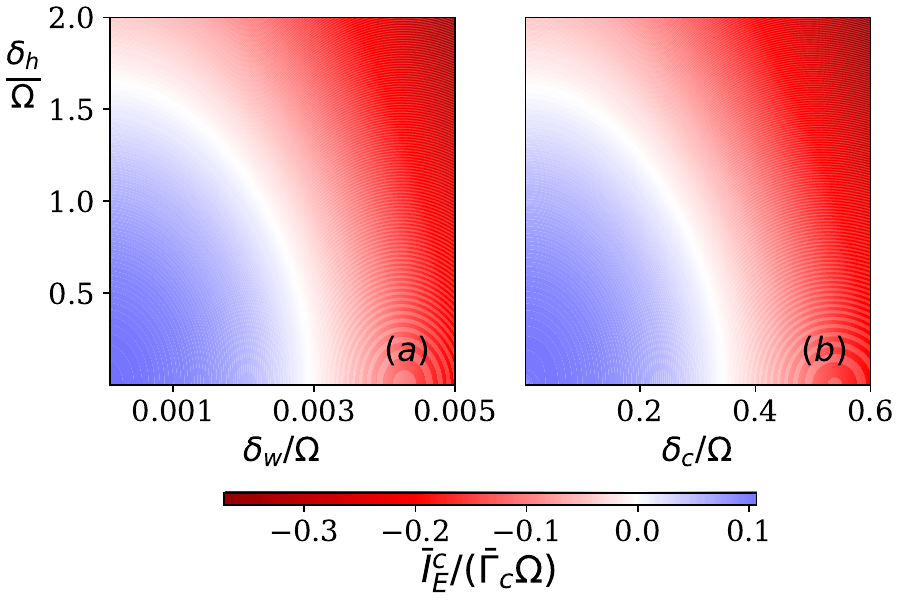}
    \caption{
    Attainability of cooling as a function of the widths $\delta_\nu$ of the spectral densities of a type~\eqref{EQ:specdens}.
    The color code shows the stationary current $\bar{I}_E^c$. 
     The blue regions denote the setting where the machine functions as a refrigerator.
     Panel (a): the cold reservoir spectral density width is fixed $\delta_c=0.1 \Omega$.
     Panel (b): the work reservoir spectral density width is fixed $\delta_w=10^{-3} \Omega$.
     The size of the system forming the WF is $N=31$.
      Other parameters are the same as in Fig.~\ref{fig:I}.}
    \label{fig:deltas}
\end{figure}
Apparently, if transitions within the WF deviate from the ideal reduced model scenario, the cooling performance gets worse.
Key factors in this are the properties of the spectral densities of the respective heat baths, namely the widths $\delta_\nu$ from the parametrization~\eqref{EQ:specdens}.
If they allow for too broad energy couplings we can expect the machine to stop working as a refrigerator as other than the desired transitions are driven~\cite{friedman2019a}.
In Fig.~\ref{fig:deltas} we identify the working region (in blue) as the function of $\delta_\nu$ for all three reservoirs.
Because the temperature of the work reservoir is supposed to be very high, the performance of our machine will be very sensitive to the ability to drive only the precise transitions in the WF.
For example, for small $\delta_w$, the high-temperature work reservoir can essentially only drive the $\ket{v_{3/2}^+}\leftrightarrow\ket{v_{5/2}^+}$ transition. 
If however $\delta_w$ is larger, the (highest-temperature) work reservoir will drive the system to other cycles (faint colours in Fig.~\ref{fig:absorbtion_refrigerator}), destroying the cooling functionality.
Indeed, in our simulations we observe that the value $\delta_w$ needs to be two or three orders smaller than $\delta_c$ and $\delta_h$ to achieve cooling.

\begin{figure}[t]
    \includegraphics[width=1\linewidth,clip=true]{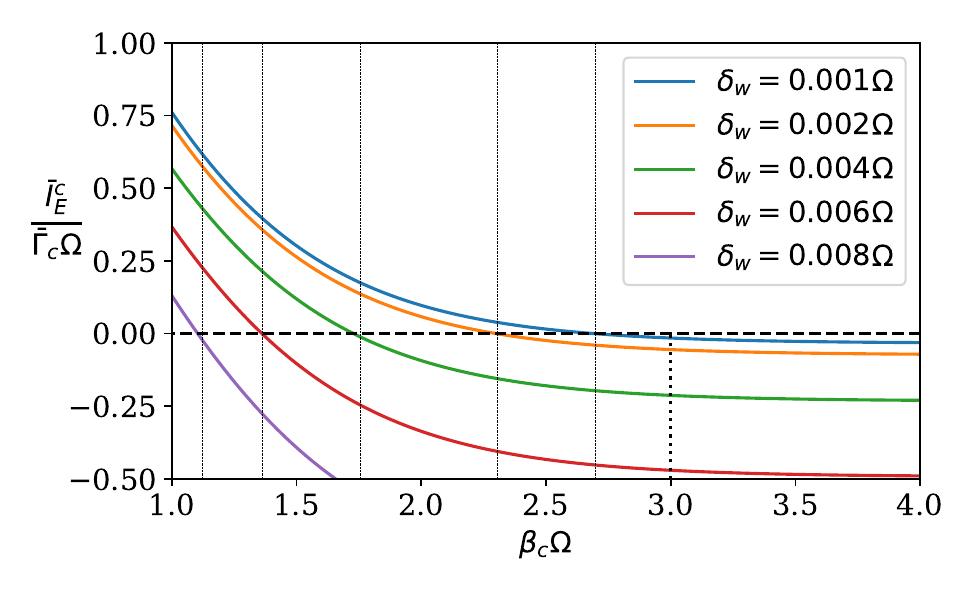}
    \caption{
    Attainability of cooling (current above the dashed horizontal line) as a function of the temperature of the cold reservoir $\beta_c$ plotted for several values of the width of the work reservoir spectral density $\delta_w$.
    The vertical lines mark at which values of $\beta_c$ the current flips its sign for each curve.
    The vertical dotted line marks the maximal inverse temperature $\beta_c^{\rm max}=3\beta_h$ at which the reduced model can operate.
    The size of the system forming the WF is $N=31$.
    Other parameters are the same as in Fig.~\ref{fig:I}.}
    \label{fig:betas}
\end{figure}
Given fixed $\delta_w$, we can also investigate what the maximal inverse temperature $\beta_c$ is, to which the machine can cool the cold reservoir.
This is depicted in Fig.~\ref{fig:betas}.
Apparently, larger $\delta_w$ results in lower values of $\beta_c^{\rm max}$ for which the  positive cooling current can be maintained.
This can be understood as the work reservoir for finite $\delta_w$ tends to excite our system to higher levels. 
There, the spectral densities are no longer tuned to the different energy differences.
Even if they were perfectly tuned e.g. to the cycle formed by the higher eigenstates $\ket{v_{5/2}^+} \to \ket{v_{7/2}^+} \to \ket{v_{9/2}^+}$, one would obtain a reduced cooling window due to the increased level spacings.
Thus, the thermal operation window becomes smaller when the spectral density of the work reservoir is not sharp.
The theoretical maximum of $\beta_c^{\rm max}=3 \beta_h$ is given by Eq.~\eqref{EQ:AnalIc} for the reduced model.
In Fig.~\ref{fig:betas} we can see that this limit is closely approachable with $\delta_w \to 0$.

From the discussion above it is clear that the precise tuning of the spectral density related to the work bath is essential as this is the most sensitive parameter.
However, as already noted, this reservoir could be implemented as a laser exclusively driving the required transition, see the discussion at the end of App~\ref{APP:reduced_model}.

\begin{figure}[ht!]
    \includegraphics[width=1\linewidth,clip=true]{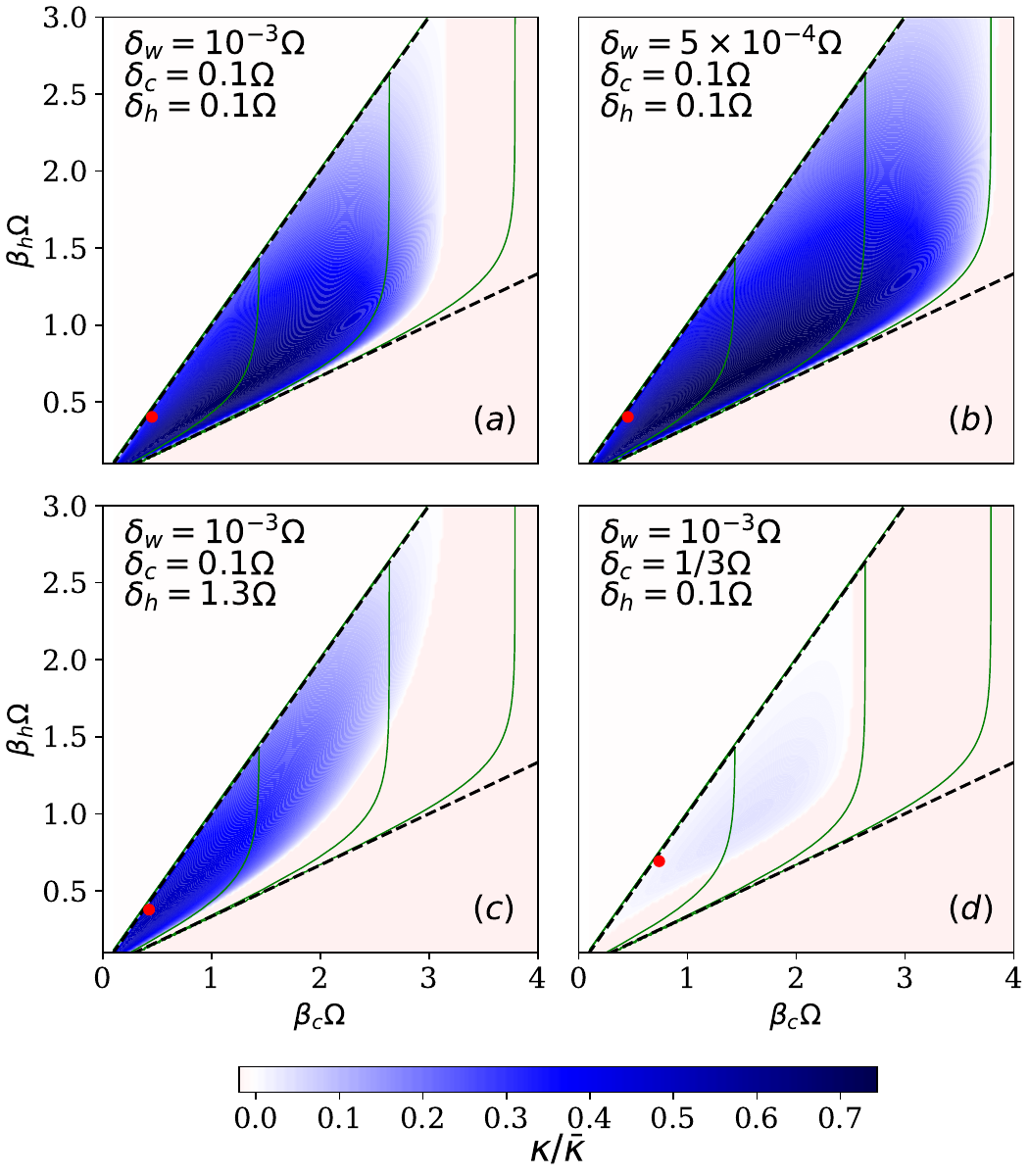}
    \caption{
    Relative coefficient of performance for cooling as a function of the inverse temperatures of the cold  $\beta_c$ and the hot $\beta_h$ reservoirs.
    The blue region marks the positive cooling current $\bar{I}_E^c>0$  and the color intensity encodes the cooling performance $\kappa/\bar{\kappa}$, see Eqns.~\eqref{EQ:CoolingEfficiency} and~\eqref{EQ:kbar}.
    The red dots indicate the highest value of $\bar{I}_E^c$.
    Panels (a)--(d) correspond to different values of $\delta_\nu$.
    The red point in each panel marks the largest relative coefficient of performance.
    The dashed lines indicate $\beta_c=\beta_h$ and $\beta_c=3 \beta_h$ which is the temperature window in which the reduced model works as a cooler, see App~\ref{APP:reduced_model}.
    The thin green solid lines in all panels mark the equivalue lines of a constant cooling current $\bar{I}_E^c$ in the case of the reduced model.
    From left to right these values are $\bar{I}_E^c/(\bar{\Gamma}_c \Omega) = 0.4, 0.04, 0.004$.
    The size of the system forming the WF is $N=31$.
    Other parameters are the same as in Fig.~\ref{fig:I}.}
    \label{fig:betas2}
\end{figure}
Now, we want to establish the operational region of the refrigerator for other parameters as well and identify the most optimal parameter setting.
To do that, in Fig~\ref{fig:betas2} we plot the cooling efficiency $\kappa/\bar{\kappa}$ [see Eqns.~\eqref{EQ:CoolingEfficiency} and~\eqref{EQ:kbar}] in the plane $\beta_c \times \beta_h$ for various values of $\delta_\nu$.
Because $\bar{\kappa}$ bears the information on the current-noise ratio, the plotted quantity provides a suitable quantifier of the efficiency of the engine in the presence of noise.
The red dots indicate the position where the cooling current is maximal for each panel.
Thus, this point represents the most optimal setting w.r.t the maximal power output for each panel.
The area between the two dashed lines marks where the reduced model functions as a cooler, see App~\ref{APP:reduced_model}.
As expected, the full model does not allow us to cool in the full range of these temperatures due to the current leakage out of the three lowest lying levels of the WF.

In Fig~\ref{fig:betas2}(a) and (b) we demonstrate again that with $\delta_w \to 0$ the cooling performance can get very close the the ideal one of the reduced model.
One can see, however, that any non-vanishing width $\delta_w$ gives rise to maximal $\beta_c$ to which the machine can cool.
Indeed, the blue region is bounded from the right which is in contrast to the reduced model where the available area between the dashed lines is unbounded in this sense.
However, already for the reduced model  one finds that for large $\beta_c$ the contours of constant cooling current proceed along vertical lines $\beta_c={\rm const}$, i.e., although for the reduced model the cooling current remains positive between the dashed lines, its value quickly decays for large $\beta_c$.
Any slight perturbation (like in our case the population leakage to the higher-energetic unproductive cycles) will thus dominate the small positive contribution from the lowest cycle, which explains the finite cooling area observed.
Indeed, for example in panel (b) the borderline between the positive and negative cooling current follows the equivalue line  $\bar{I}_E^c = 0.004 \ \bar{\Gamma}_c \Omega$ of the reduced model.
Panels~(c) and~(d) show how the operational diagram changes with larger $\delta_c$ and $\delta_h$.
Generally, cooling at $\beta_c$ requires larger values of $\beta_h$ than in the reduced model.
The cooling efficiency is also generally smaller as compared to panels (a) and (b) which is especially the case in panel (d) where the also the relative efficiency is very low overall. 
Thus, the cooler is more sensitive to $\delta_c$ than $\delta_h$ which agrees with our findings in Fig.~\ref{fig:deltas}(b).

\section{Summary}\label{SEC:summary}

We have suggested a model for interacting two-level systems that -- when dissipatively driven by three independent reservoirs -- perform a useful function as a collective quantum absorption refrigerator. 
Globally valid conservation laws allowed us to perform a simplified analysis within the standard framework of rate equations.
Despite this fact, we could find a true quadratic quantum speedup in the cooling current, while standard thermodynamic laws were obeyed.
We also remark that the boosted cooling current cannot break the third law of thermodynamics, since the operational cooling window is always finite.

In suitable parameter regimes -- narrow spectral densities driving desired transitions exclusively and furthermore not too large temperatures -- we could further reduce the state space of our model to an ideal quantum absorption refrigerator, for which analytic results are available. 
The full counting statistics formalism allowed us to investigate deviations of our model from this ideal performance.
Here, we found a reduced cooling operational window, which however did not affect the quadratic speedup.

We emphasize that -- despite the simple rate equation structure of our underlying equations -- our results are not in conflict with Refs.~\cite{holubec2019,liu2021a}, as coherence is a basis-dependent concept. 
Indeed, our \changes{proper collective} basis features highly coherent states in the local basis of the individual working fluid particles.

We did not consider corrections to the collectivity of our model nor to the parity conservation \changes{explicitly}.
If the initial preparation prepares a small fraction of the population in other subspaces, we do not expect significant corrections as the subspaces still evolve independently, \changes{see App. \ref{APP:fragility}}.
However, imperfect implementations of our Hamiltonian may lead to always-present transitions between the subspaces, for which we expect a quick breakdown of the cooling function -- just as it would not work for $N$ independently driven two-level systems.
\changes{Likewise, we did not consider the effect of stronger system-reservoir coupling strengths. 
While one can of course expect corrections~\cite{bhandari2021a} for stronger couplings, we do not expect these to directly affect the scaling behaviour of the refrigerator, since for a perfectly implemented Hamiltonian the underlying symmetries are respected to all orders.
To model such imperfections} would be a technically more demanding challenge but as our model can be implemented with at least $N=5$ two-level systems, corresponding studies are within reach and are an interesting possibility of further research. 

\changes{
Finally, we remark that although we have treated the cold reservoir by non-interacting bosons, the true benefit of a boosted cooling functionality would be revealed if that reservoir is replaced by a quantum system of interest.
This would enable one to explore the various quantum phenomena near ground states of interacting many-body systems, or to find solutions to quantum search problems encoded in such ground states.
We therefore believe that our first-step exposition can be useful to design machines exploiting quantum effects.
}

\section*{Acknowledgements}

The authors acknowledge the discussion with Martin Koppenh\"ofer.
M.K. was financially supported by the Swiss National Science Foundation (SNSF) and the NCCR Quantum Science and Technology.
G.S. acknowledges support by the Helmholtz high-potential program and previous support by the DFG.

\appendix

\changes{
\section{Fragility considerations}\label{APP:fragility}

As discussed in the main text, our model operates at steady state in the maximum angular momentum and positive parity subspace, and thus requires the initial preparation of some representative state of that subspace.
For example, the energetically highest state of that sector $\ket{v_{N/2}^+} = \frac{1}{\sqrt{2}} \left[\ket{N/2, +N/2}+\ket{N/2,-N/2}\right] = \frac{1}{\sqrt{2}}\left[\ket{0\ldots 0}+\ket{1 \ldots 1}\right]$ is a Greenberger-Horne-Zeilinger (GHZ) state~\cite{greenberger1990a}.
The $N$-qubit GHZ state could be prepared by unitary pulses from a product state via $e^{\pm \ii \pi/2 \sigma^1_y \otimes \ldots \otimes \sigma^N_y} \ket{0\ldots 0}$ (where the sign depends on whether $(N+1)/2$ is even or odd).
Also measurements of the parity would prepare the GHZ state from a simple product state.
Alternatively, one could relax to the ground state of a finite-size Lipkin-Meshkov-Glick model in the normal phase and then adiabatically deform it~\cite{Klo19} into the superradiant phase -- where the Hamiltonian corresponds to our Eq.~\eqref{EQ:Hs} and has the even parity ground state $\ket{v_{1/2}^+}$.
Such schemes all appear challenging to implement but appear within reach of current experimental abilities, as recent experiments~\cite{zhao2021a} have created GHZ states on thousands of qubits.

We consider the consequences of imperfect initial preparation below, which suggest that for odd $N$ imperfect initial preparation is not the main error source.

\subsection{Imperfect angular momentum preparation}

If instead one prepares a fraction of the system's population in other subspaces with smaller total angular momentum (but even parity), the structure of our system Hamiltonian~\eqref{EQ:Hs} would not change, but $m$ could only assume smaller values.
That is, the relevant lower states in Fig.~\ref{fig:absorbtion_refrigerator} would remain unchanged, but the Clebsch-Gordan coefficients~\eqref{EQ:Clebsch} (and derived transition rates) would be reduced by inserting smaller $j$ values.
If for large $N$ the current of the maximum-$j$ sector ($j=N/2$) scales as 
$I_{j} \approx \alpha j^2$, the average cooling current (with $P_{j}, P_{j-1}, \ldots$ denoting the probabilities of the respective sectors in the initial state preparation) would be
\begin{align}
    I \approx P_{j} I_{j} + P_{j-1} I_{j-1} + \ldots\,,
\end{align}
which is not a drastic modification as long as only the large-$j$ sectors are occupied.

\subsection{Imperfect parity preparation}

If we assume that we have prepared a mixture of even and odd parity sectors, things depend on whether $N$ is even or odd.

If $N$ is odd, both even and odd parity sectors are completely symmetric (this then also holds for the other angular momentum subspaces).
Therefore, all transitions are the same, and both sectors would contribute equally to the cooling (of course only in the operational window).
The functionality would in this case not be diminished at all.

In contrast, if $N$ is even, the even parity sector (with states $\{\ket{v_0^+},\ket{v_1^+}, \ket{v_2^+},\ldots\}$) has different energy differences than the odd parity sector
(with states $\ket{v_1^-}, \ket{v_2^-}, \ket{v_3^-},\ldots\})$~\footnotemark[2]. 
Thus, if the even sector is tuned to operate as absorbtion refrigerator, the other sector will most certainly be out of its operational window.
In the best case (e.g. if the spectral densities are highly peaked on the  $\Omega$, $3\Omega$, and $4\Omega$ transitions that exist in the even parity sector for even $N$), it would simply mean that the odd sector does not contribute to steady state cooling, as there only the $3\Omega$-transition of the work reservoir could drive the transitions $\ket{v_1^-} \leftrightarrow \ket{v_2^-}$.
In the worst case (e.g. broader spectral densities), the cycle of the odd subspace could also be traversed in the opposite direction, effectively  superradiantly heating the cold reservoir, such that this case should be avoided.
}

\section{Generating peaked spectral densities via an inverted reaction-coordinate mapping}\label{APP:specdens_mapping}

\begin{figure}
    \centering
    \includegraphics[width=0.45\textwidth,clip=true]{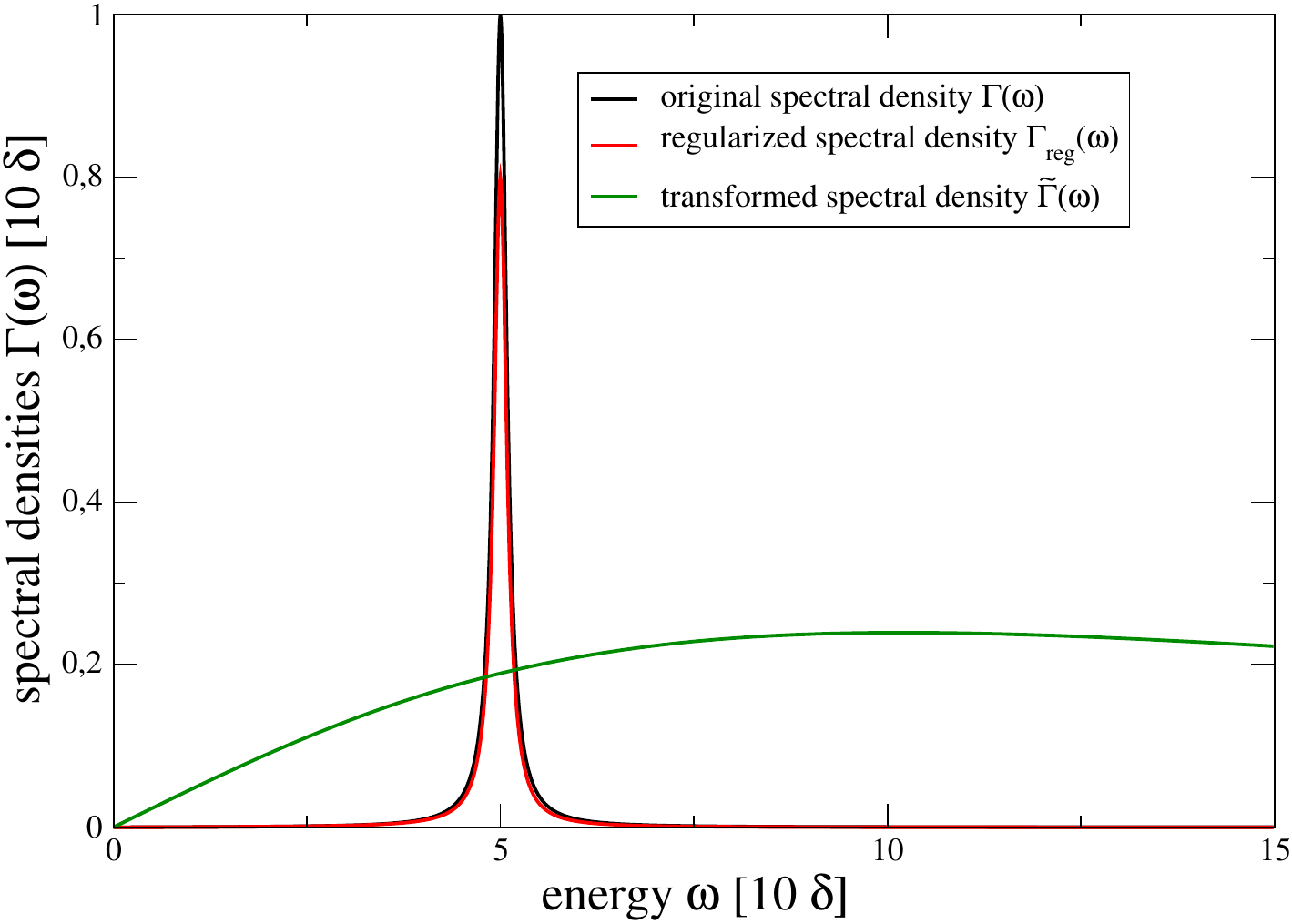}
    \caption{Mapping procedure for spectral densities. Since the original spectral density (black) does not decay fast enough, we consider the regularized version (red). The mapped spectral density (green) is of Lorentz-Drude type and rather structureless.} 
    \label{fig:rcmapping}
\end{figure}

Harmonic reservoirs may be reorganized by Bogoliubov transforms, and if these reservoirs are coupled by spectral density $\Gamma_\nu(\omega)$ to a system, such transformations will also affect the spectral density.
A particularly important mapping -- the reaction-coordinate mapping -- reorganizes the Hamiltonian such that the original system is first coupled with coupling strength $\tilde{\lambda}_\nu$ to a single collective bosonic mode of energy $\tilde{\Omega}_\nu$ that is then coupled to a residual reservoir via a transformed spectral density $\tilde{\Gamma}_\nu(\omega)$.
Specifically, these parameters are fully determined by the original spectral density of the model (see e.g. Eqns~(5-7) of Ref.~\cite{nazir2019a}).
The squared reaction coordinate energy is given by
\begin{align}\label{EQ:rcenergy}
    \tilde{\Omega}^2 = \frac{\int_0^\infty \omega^3 \Gamma(\omega) d\omega}{\int_0^\infty \omega \Gamma(\omega) d\omega}\,,
\end{align}
and the squared coupling strength between system and reaction coordinate computes as
\begin{align}
    \tilde{\lambda}^2 = \frac{1}{2\pi\tilde{\Omega}}\int_0^\infty \omega \Gamma(\omega) d\omega\,.
\end{align}
Then, the mapped residual spectral density is obtained via
\begin{align}
    \tilde{\Gamma}(\omega) = \frac{4\tilde{\lambda}^2\Gamma(\omega)}{\left[\frac{1}{\pi} {\cal P} \int\limits_{-\infty}^{+\infty} \frac{\Gamma(\omega')}{\omega'-\omega}\right]^2+\left[\Gamma(\omega)\right]^2}\,,
\end{align}
where ${\cal P}$ denotes the principal value.
From this we see that scaling the original spectral density by a constant leaves the mapped spectral density invariant, and this argument is typically used to motivate a perturbative treatment for the supersystem composed of original system and reaction-coordinate even in the limit of strong coupling with large $\Gamma(\omega)$.
Unfortunately, the first numerator integral in~\eqref{EQ:rcenergy} does not converge when we try this for our assumed spectral density~\eqref{EQ:specdens} in the main text.
Therefore, we consider a regularized spectral density
\begin{align}\label{EQ:specdens_reg}
    \Gamma_{\rm reg}(\omega) = \Gamma(\omega) \frac{\Delta^2}{\Delta^2+\omega^2}\,,
\end{align}
with $\Gamma(\omega)$ taken from~\eqref{EQ:specdens}  (for brevity we drop the index $\nu$ since it is individually done for every reservoir).
Then we get
\begin{align}
    \tilde{\Omega}^2 &= \varepsilon^2 + \delta^2 + 2 \Delta \delta\,,\qquad
    \tilde{\lambda}^2 = \frac{1}{2\pi\tilde{\Omega}} \frac{\Gamma \pi \delta \varepsilon \Delta^2}{\varepsilon^2+(\Delta+\delta)^2}\nn
    \tilde{\Gamma}(\omega) &= 2\frac{\delta}{\tilde{\Omega}} \left[\varepsilon^2+(\Delta+\delta)^2\right] \frac{\omega}{\omega^2+(\Delta+2\delta)^2}\,,
\end{align}
which is a Lorentz-Drude spectral density, see Fig.~\ref{fig:rcmapping}.
Since Bogoliubov transforms are invertible, we can also turn the perspective.
Then, a system that is via coupling $\tilde{\lambda}$, coupled to a harmonic mode of energy $\tilde{\Omega}$ that is further coupled to a reservoir with a Lorentz-Drude spectral density, feels the presence of the harmonic mode with its reservoir as an effectively highly peaked spectral density~\eqref{EQ:specdens_reg}, which for $\Delta\to\infty$ reduces to~\eqref{EQ:specdens} in the main text.
Put more generally, the harmonic mode acts as an energy filter between system and reservoir that generates a sharply peaked spectral density seen by the system.

\section{Superradiant thermalization close to the ground state for one reservoir} \label{APP:SuperradThermalization}

Connecting the WF of a type~\eqref{EQ:Hs} to a single heat bath with the dipole operator $J_x$ gives rise to the population dynamics, i.e. Eq.~\eqref{EQ:RateEquation}, which thermalizes the system in the basis~\eqref{EQ:vs}.
The explicit dynamics of the population in the state $\ket{v_a^+}$ is given as
\begin{equation}
\begin{split}
\dot{\rho}_a=&\frac{1}{4}\gamma(E_{\abs{a-1}}-E_a)\abs{\alpha^+_{a-1}}^2\rho_{\abs{a-1}}\\
&+\frac{1}{4}\gamma(E_{a+1}-E_a)\abs{\alpha^-_{a+1}}^2\rho_{a+1}\\
&-\frac{1}{4}\gamma(E_{a}-E_{a+1})\abs{\alpha^+_a}^2\rho_a\\
&-\frac{1}{4}\gamma(E_a-E_{\abs{a-1}})\abs{\alpha^-_{a}}^2 \rho_a\,.
\end{split}
\label{EQ:diagonalsEvo_app}
\end{equation}
Here $\gamma$ is the energy-dependent coupling strength from Eq.~\eqref{EQ:small_gamma}.
Note that $\gamma(-\omega)=\Gamma(\omega)n(\omega)$ where $\Gamma(\omega)$ is a spectral density of the reservoir and $n(\omega)$ the Bose-Einstein distribution.
The Clebsch-Gordan coefficients  $\alpha_a^{\pm}$ defined in Eq.~\eqref{EQ:Clebsch} are responsible for the $N^2$ boost in the dynamics close to the ground state $a \ll N$.
They fulfill a trivial identity $\alpha^+_a=\alpha^-_{-a}$.
Equation.~\eqref{EQ:diagonalsEvo_app} takes the same form for $N$ odd and even.

To quantify the thermalization time $t_{\rm th}$ we compute the time-dependent relative entropy $ S(\rho(t)||\rho_\beta)= \trace{\rho(t)\left( \log{\rho(t)}-\log{\rho_\beta} \right)}$ between the state $\rho(t)$ and the thermal state $\rho_\beta$ given by the inverse temperature $\beta$ of the reservoir.
We define $t_{\rm th}$ as the time point where $S(\rho(t)||\rho_\beta) \leq 10^{-6}$. Because of the contractivity of Markovian dynamics such definition of $t_{\rm th}$ is unique.

\begin{figure}[ht!]
    \includegraphics[width=1\linewidth,clip=true]{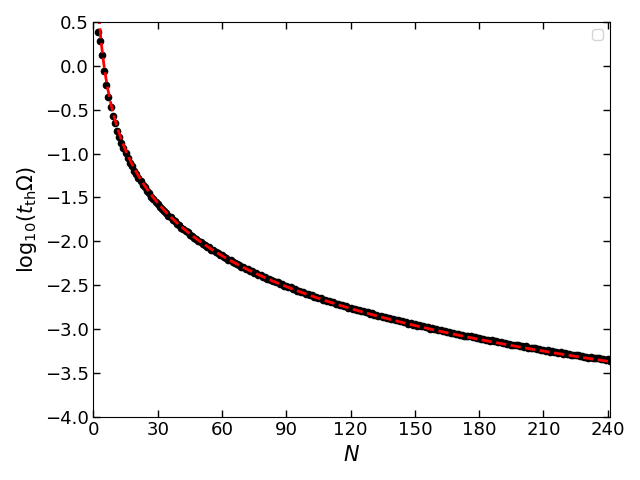}
    \caption{Thermalization time $t_{\rm th}$ as a function of the size of the WF given by the Hamiltonian~\eqref{EQ:Hs}.
    The black points are the numerical values computed using Eq.~\eqref{EQ:diagonalsEvo_app} for integer $N$ and the red dashed line is $1/N^2$ fit. 
    The initial state is a thermal state with $\beta_{\rm i} = 1 \Omega^{-1}$.
    The temperature of the heat bath is $\beta_{\rm f} = 4 \Omega^{-1}$.}
    \label{fig:ThermTime}
\end{figure}

In the following we use the Ohmic spectral density of the heat reservoir $\Gamma(\omega)= \omega e^{-\omega/\omega_c}$ with the cut-off $\omega_c = 100 \Omega$  and we set its (inverse) temperature to $\beta_{\rm f} = 4 \Omega^{-1}$.
Initiating the WF in the thermal state with the inverse temperature $\beta_{\rm i} = 1 \Omega^{-1}$ we compute the thermalization time $t_{\rm th}$ using Eq.~\eqref{EQ:diagonalsEvo_app} after bringing the WF into contact with the heat bath, see Fig.~\ref{fig:ThermTime}.  
Apparently, for growing $N$, the fixed temperatures $\beta_{\rm i}$ and $\beta_{\rm f}$ allow to populate only the low lying levels of the WF.
In our model of the WF these are exactly the levels whose dynamics is boosted by the Clebsch-Gordan.
Thus, we observe $1/N^2$ scaling of the thermalization time $t_{\rm th}$.
This is in contrast to the system on $N$ mutually non-interacting spins forming the WF which are collectively coupled to the heat reservoir~\cite{Klo19}.
Such a system, described by a collective Hamiltonian $H=\Omega J_z$, can also exhibit superradiant scaling in the maximal $j$ subspace, however only when the dynamics takes place in the middle part of the spectrum $m \approx 0$ where $m$ labels the $J_z$ eigenbasis.
If the dynamics is restricted to the region around the ground state only, the superradiant boost is lost.
Indeed, the thermalization time $t_{\rm th}$ would scale simply as $1/N$, cf. Fig. 5 in Ref.~\cite{Klo19}.

In the zero-temperature limit $\beta_{\rm f} \to \infty$, an even simpler analysis of the relaxation process is possible by means of waiting-time distributions.
Then, the transition rates from state $\ket{v_{a+1}^+}$ to state $\ket{v_a^+}$ are just given by $R_{a,a+1}=\Gamma(E_{a+1}-E_a)/4\left[\frac{N}{2} \left(\frac{N}{2}+1\right)-a(a+1)\right]$, and when splitting the generalized rate matrix as $R(\chi) = R_0 + R_1(\chi)$ with the diagonal matrix $R_0$ containing the no-jump evolution, we can easily construct the waiting-time distribution $\omega(\tau)$ from the time-derivative of the probability for no jump $P_0(\tau)$.
This yields
\begin{align}
    \omega(\tau) = -\frac{d}{d\tau} P_0(\tau) = -\trace{R_0 e^{R_0\tau}\rho_0}\,,
\end{align}
which can be simply evaluated as $R_0$ is a diagonal matrix.
For example, when $\rho_0=\ket{v_{a+1}^+}\bra{v_{a+1}^+}$ is also diagonal, the waiting time distribution for the decay of state $\ket{v_{a+1}^+}$ into $\ket{v_a^+}$ is just
$\omega_{a,a+1}(\tau) = R_{a,a+1} e^{-R_{a,a+1} \tau}$ with average waiting time $\expval{\tau_{a,a+1}}=[R_{a,a+1}]^{-1}$.
Since at zero temperature, relaxation to the ground state is the only decay channel, we may just add the individual waiting times to get an estimate on the total relaxation time.
For example, initializing the system in the state $\ket{v_{a_0}^+}$ with $a_0\ge 3/2$, we get 
\begin{align}
    \expval{\tau} &= \sum_{a=3/2}^{m_0} \expval{\tau_{a-1,a}}\nn
    &= \sum_{a=3/2}^{a_0} \frac{4}{\Gamma((2a-1)\Omega) \left[\frac{N}{2}\left(\frac{N}{2}+1\right)-a(a-1)\right]}\,.
\end{align}
When $a_0$ is not much larger than $3/2$, the above sum will only contain few terms, and the $1/N^2$ scaling of the relaxation time becomes obvious.
When $a_0=N/2$, the relaxation dynamics may be modified by the behaviour of the spectral density but may for roughly constant $\Gamma(\omega)$ still scale as $1/N$.

\section{Supertransmittance close to the ground state for two reservoirs}
\label{APP:supertransmittance}

If our system is only coupled to work and cold reservoir ($\bar\Gamma_h=0$) and the spectral densities of the remaining cold and work reservoir are not highly peaked, we generate a collective transport situation similar to Ref.~\cite{vogl2011a}, but with the difference that there the system was just a collection of non-interacting spins.
In that system, due to the equidistant spectrum, an analytic calculation of the nonequilibrium steady state was possible in terms of some average temperature. 
There, collective transport enhancements were only visible in regimes where both reservoirs were rather hot, essentially because large Clebsch-Gordan coefficients  were encountered only in the middle of the spectrum.
In contrast, for our system the nonequilibrium steady state is not so easy to obtain in a closed analytic form as spectrum of $H_S=\Omega J_z^2$ is not equidistant. 
However, when both reservoirs are at low temperature $\beta_{c/w} 2\Omega \gtrapprox 1$, we also find a quadratically scaled current (from work reservoir to cold reservoir $\abs{\bar{I}_E^c}\propto N^2$), which can be understood analytically.
If there is no hot reservoir and work and cold reservoir act analogously, the red and green transitions in Fig.~\ref{fig:absorbtion_refrigerator} should be superimposed and drive only transitions between neighbouring states $\ket{v_a^+} \leftrightarrow \ket{v_{a\pm 1}^+}$.
Moreover, if already $\beta_{c/w}2\Omega \gtrapprox 1$, this is even more fulfilled for the higher-energy transitions (shaded in Fig.~\ref{fig:absorbtion_refrigerator}), and it will be much more likely to relax the system than to excite it.
As a result, the dynamics is dominantly described by the two-lowest states
$\{\ket{v_{1/2}^+},\ket{v_{3/2}^+}\}$, where we have the rate matrix
$R(\chi)=R_c(\chi)+R_w$ with
\begin{align}
    R_c(\chi) &= \Gamma_c^{\rm eff} \left(\begin{array}{cc}
    -n_c & +(1+n_c) e^{-\ii\chi 2\Omega}\\
    +n_c e^{+\ii\chi 2\Omega} & -(1+n_c)
    \end{array}\right)\,,\nn
    R_w &= \Gamma_w^{\rm eff} \left(\begin{array}{cc}
    -n_w & +(1+n_w)\\
    +n_w & -(1+n_w)
    \end{array}\right)\,,
\end{align}
where $n_c = \frac{1}{e^{\beta_c 2\Omega}-1}$ and 
$n_w=\frac{1}{e^{\beta_w 2\Omega}-1}$ denote the Bose distributions evaluated at  the lowest energy difference $E_{3/2}-E_{1/2}=2\Omega$ and where
$\Gamma_{c/w}^{\rm eff} = \Gamma_{c/w}(2\Omega) \abs{\bra{v_{1/2}^+}J_x\ket{v_{3/2}^+}}^2 = \Gamma_{c/w}(2\Omega)/4 \left[\frac{N}{2}\left(\frac{N}{2}+1\right)-\frac{3}{4}\right]$ 
are the effective bare transition rates determined by spectral densities~\eqref{EQ:specdens} and the Clebsch-Gordan coefficients~\eqref{EQ:Clebsch}.
As both $\Gamma_{c/w}^{\rm eff}\propto N^2$, the dominant eigenvalue $\lambda(\chi)$ and therefore all cumulants~\cite{touchette2009a} will scale quadratically with $N$ in this regime.
From its first derivative with respect to the counting field $\bar{I}_E^c = -\ii \partial_\chi \lambda(\chi)|_{\chi=0}$ (or, alternatively by computing it via~\eqref{EQ:energy_current}), we get the current from the cold reservoir
\begin{align}
    \bar{I}_E^c \to -\frac{2\Omega \Gamma_c^{\rm eff} \Gamma_w^{\rm eff} (n_w-n_c)}{\Gamma_c^{\rm eff} [1+2 n_c]+ \Gamma_w^{\rm eff}[1+2 n_w]}\,,
\end{align}
which is negative when $n_w>n_c$ (as heat flows from the work reservoir through the system to the cold reservoir) and changes sign under reversing the thermal biases $\beta_c\leftrightarrow\beta_w$.
Analogous to known results for bosonic transport through a two-level system~\cite{segal2005b}, the above formula may display some mild rectification effects for asymmetric coupling strengths -- with the difference that the effective coupling constants scale quadratically with $N$.

In contrast, considering the limit $\bar\Gamma_w=0$ and $\delta_h\to 0$ with $\varepsilon_h=6\Omega$, the two remaining reservoirs act very differently and one can observe stronger rectification effects under exchanging the thermal bias.
However, the rectification ratio is not further enhanced with increasing $N$.

\section{Current and noise formulas}\label{APP:current_noise}

The moment-generating function for the distribution of energies entering the system from the cold reservoir during the interval $[0,t]$ is given by
\begin{align}
    M(\chi,t) = \trace{\rho(\chi,t)} = \trace{e^{R(\chi) t} \rho_0}\,.
\end{align}
From it, the moments of the energies transferred can be obtained by taking derivatives $\expval{E^k}_t = (-\ii \partial_\chi)^k M(\chi,t)|_{\chi=0}$.

To evaluate the energy current from the cold reservoir, we compute the time-derivative of the first moment
\begin{align}
    I_E(t) &= \frac{d}{dt} \expval{E}_t = (-\ii\partial_\chi) \trace{R(\chi) e^{R(\chi) t} \rho_0}|_{\chi=0}\nn
    &= -\ii \trace{R'(0) \rho(t)}\,,
\end{align}
where we have used that $\trace{R(0) v}=0$ for any vector $v$ due to the trace-preservation of the rate matrix $R(0)$ and furthermore $\rho(t) = e^{R(0) t} \rho_0$.
In the long-term limit this simplifies to~\eqref{EQ:energy_current} in the main text, where $\bar\rho$ is just the -- properly normalized -- steady state solution of the rate equation that is defined by Eq.~
\eqref{EQ:steady_state}.

The noise is given by the time-derivative of the second cumulant.
Since we cannot analytically obtain the dominant eigenvalue of the resulting $(N+1)/2\times(N+1)/2$-dimensional rate matrix $R(\chi)$, we obtain the noise by adapting the methods developed for the counting-statistics of time-dependent (driven) conductors~\cite{benito2016b,restrepo2019a} to the simpler undriven case.
The time-dependent noise then becomes
\begin{align}\label{EQ:noise_app}
    S_E(t) &= \frac{d}{dt} \left[\expval{E^2}_t - \expval{E}_t^2\right]\nn
    &= (-\ii \partial_\chi)^2 \trace{R(\chi) e^{R(\chi) t}\rho_0} -2 I_E(t) \expval{E}_t\nn
    &= -\trace{R''(0) \rho(t)} -2\trace{R'(0) \left(\partial_\chi e^{R(\chi) t}|_{\chi=0}\right)\rho_0}\nn
    &\qquad+ 2 \trace{R'(0) \rho(t)} \trace{\left(\partial_\chi e^{R(\chi) t}|_{\chi=0}\right) \rho_0}\nn
    &= -\trace{R''(0)\rho(t)} -2 \trace{R'(0) \sigma(t)}\,,
\end{align}
where we have again used the trace conservation property of the rate matrix $R(0)$ and in the last step we have combined the last two terms by introducing
the auxiliary quantity (that we arranged as a vector as well)
\begin{align}
    \sigma(t) &= \partial_\chi \left.\frac{e^{R(\chi) t} \rho_0}{\trace{e^{R(\chi) t} \rho_0}}\right|_{\chi=0}\\
    &=\left(\partial_\chi e^{R(\chi)t}|_{\chi=0}\right)\rho_0-\rho(t) \trace{\left(\partial_\chi e^{R(\chi)t}|_{\chi=0}\right)\rho_0}
    \,.\nonumber
\end{align}
By construction, we have $\trace{\sigma(t)}=(1,\ldots,1)\cdot \sigma=0$.
Furthermore, the auxiliary obeys the differential equation
\begin{align}
    \dot{\sigma} = R'(0)\rho(t) + R(0) \sigma(t) - \rho(t) \trace{R'(0) \rho(t)}\,,
\end{align}
which -- to obtain the full time-dependent noise -- would have to be solved with the initial condition $\sigma(0)=(0,\ldots,0)^{\rm T}$.
However, if we are just interested in the steady-state values, we may set the r.h.s. of the above equation to zero and insert $\rho(t)\to\bar\rho$ to solve for $\bar\sigma$, leading to~\eqref{EQ:steady_sigma} in the main text.
Equation~\eqref{EQ:noise_app} then reduces in the long-term limit to Eq.~\eqref{EQ:energy_noise} in the main text.

\changes{
The generalization of such methods to non-Markovian systems~\cite{flindt2008a,flindt2010a} or time-dependently driven ones~\cite{benito2016b,restrepo2019a} is also possible.
}

\section{Bound for the coefficient of performance for three reservoirs}
\label{APP:carnotbound}

In the regime where $\beta_c>\beta_h>\beta_w$ it follows from Eqns.~\eqref{EQ:first_law} and~\eqref{EQ:second_law} that cooling of the coldest reservoir ($\bar{I}_E^c>0$) is only possible via heat entering from the work reservoir ($\bar{I}_E^w>0$) and dumping the waste heat into the hot reservoir ($\bar{I}_E^h<0$).
From the positivity of the entropy production rate~\eqref{EQ:second_law} one can then conclude that the cooling coefficient of performance (we now explicitly assume that $\beta_c \ge \beta_h \ge \beta_w$ and $\bar{I}_E^c\ge 0$ and $\bar{I}_E^w\ge 0$)
\begin{align}
    \kappa &\equiv \frac{\bar{I}_E^c}{\bar{I}_E^w}\nn
    &= \frac{\beta_h \bar{I}_E^c}{\left[-(\beta_c \- \beta_w)\bar{I}_E^c\- (\beta_h \- \beta_w)\bar{I}_E^h \right]\+\beta_w \bar{I}_E^w\+(\beta_c\-\beta_h)\bar{I}_E^c}\nn
    &\le \frac{\beta_h \bar{I}_E^c}{\beta_w \bar{I}_E^w+(\beta_c-\beta_h)\bar{I}_E^c}
    \le \frac{\beta_h}{\beta_c-\beta_h} \equiv \kappa_{\rm Ca}
    \label{EQ:CoolingEfficiency}
\end{align}
is upper-bounded by the usual Carnot bound.
The second line above can be verified by using energy conservation~\eqref{EQ:first_law}.
We have written it in a way that all three summands of the denominator are separately positive -- positivity of the term in square brackets is seen from combining~\eqref{EQ:second_law} and~\eqref{EQ:first_law} -- and the inequalities then follow by discarding parts of the denominator.

Analogously, we can use Eq.~\eqref{EQ:turbound} to obtain Eq.~\eqref{EQ:kbar} (under the same conditions as above) as follows
\begin{align}
    \kappa &=    
    \frac{\beta_h \bar{I}_E^c}{\bar\sigma_\ii + \beta_w \bar{I}_E^w+(\beta_c-\beta_h)\bar{I}_E^c}\nn
    &\le \frac{\beta_h \bar{I}_E^c}{\bar\sigma_\ii + (\beta_c-\beta_h)\bar{I}_E^c}\nn
    &\le \frac{\beta_h \bar{I}_E^c}{\frac{2 (\bar{I}_E^c)^2}{\bar{S}_E^c} + (\beta_c-\beta_h)\bar{I}_E^c}\nn
    &= \kappa_{\rm Ca} \frac{1}{1+2\frac{\bar{I}_E^c}{\bar{S}_E^c(\beta_c-\beta_h)}}\,,
\end{align}
where in the second inequality we have used the TUR~\eqref{EQ:turbound}.

\section{Reduced model}\label{APP:reduced_model}

In the ideal limit where our system is described by the lowest states $\{\ket{v_{1/2}^+},\ket{v_{3/2}^+},\ket{v_{5/2}^+}\}$ only with the three transitions driven exclusively by the reservoirs, the rate matrix assumes the form
$R(\chi)=R_c(\chi)+R_h+R_w$ with
\begin{align}
    R_c(\chi) &= \Gamma_c^{\rm eff} \left(\begin{array}{ccc}
    -n_c & +(1+n_c) e^{-\ii\chi 2\Omega} & 0\\
    +n_c e^{+\ii\chi 2\Omega} & -(1+n_c) & 0\\
    0 & 0 & 0
    \end{array}\right)\,,\nn
    R_h &= \Gamma_h^{\rm eff} \left(\begin{array}{ccc}
    -n_h & 0 & +(1+n_h)\\
    0 & 0 & 0\\
    +n_h & 0 & -(1+n_h)
    \end{array}\right)\,,\nn 
    R_w &= \Gamma_w^{\rm eff} \left(\begin{array}{ccc}
    0 & 0 & 0\\
    0 & -n_w & +(1+n_w)\\
    0 & +n_w & -(1+n_w)
    \end{array}\right)\,,
\end{align}
where $n_c = \frac{1}{e^{\beta_c 2\Omega}-1}$, $n_h=\frac{1}{e^{\beta_h 6\Omega}-1}$, and 
$n_w=\frac{1}{e^{\beta_w 4\Omega}-1}$ denote the Bose distributions evaluated at the respective energy differences.
Technically, this limit is achieved with perfectly tuned  $\varepsilon_c=2\Omega$, $\varepsilon_h=6\Omega$, $\varepsilon_w=4\Omega$ and highly peaked spectral densities $\Omega \gg \delta_\nu \to 0$ to drive the transitions exclusively.
A system that is initially prepared in one of the three lowest state will then not be able to leave it.
Within this subspace, the effective bare transition rates become
\begin{align}
\Gamma_c^{\rm eff} &= \bar{\Gamma}_c \abs{\matr{v_{1/2}^+}{J_x}{v_{3/2}^+}}^2 = \frac{\bar\Gamma_c}{4}\left(\frac{N}{2}\left(\frac{N}{2}-1\right)-\frac{3}{4}\right)\,,\nn
\Gamma_w^{\rm eff} &= \bar{\Gamma}_w \abs{\matr{v_{3/2}^+}{J_x}{\ket{v_{5/2}^+}}}^2 = 
\frac{\bar\Gamma_w}{4}\left(\frac{N}{2}\left(\frac{N}{2}-1\right)-\frac{15}{4}\right)\,,\nn
\Gamma_h^{\rm eff} &= \bar{\Gamma}_h \abs{\matr{v_{1/2}^+}{\frac{J_x^2}{N}}{v_{5/2}^+}}^2\\
&= \frac{\bar\Gamma_h}{16 N^2} 
\left(\frac{N}{2}\left(\frac{N}{2}-1\right)-\frac{3}{4}\right)\left(\frac{N}{2}\left(\frac{N}{2}-1\right)-\frac{15}{4}\right)\,,\nonumber
\end{align}
which makes their quadratic scaling with $N$ very apparent.

With standard methods (see the previous section or Ref~\cite{segal2018a}) one can calculate current and noise.

The analysis becomes particularly simple in the limit  $n_w\to\infty$, where one can coarse-grain the two excited states into one.
For the coarse-grained probabilities $P_0=P_{1/2}^+$ and $P_{12}=P_{3/2}^++P_{5/2}^+$ we can -- by using that the conditional probabilities $P_{3/2}^+/P_{12}$ and $P_{1/2}^+/P_{12}$ both approach $1/2$ -- set up a simpler rate matrix
\begin{align}
    \tilde{R}(\chi)&=\left(\begin{array}{cc}
    -\Gamma_c^{\rm eff} n_c & \frac{\Gamma_c^{\rm eff}}{2} (1+n_c)e^{-\ii\chi 2\Omega}\\
    +\Gamma_c^{\rm eff} n_c e^{+\ii\chi 2\Omega} & -\frac{\Gamma_c^{\rm eff}}{2} (1+n_c)
    \end{array}\right)\nn
    &\qquad+\left(\begin{array}{cc}
    - \Gamma_h^{\rm eff} n_h & +\frac{\Gamma_h^{\rm eff}}{2}(1+n_h)\\
    + \Gamma_h^{\rm eff} n_h & -\frac{\Gamma_h^{\rm eff}}{2} (1+n_h)
    \end{array}\right)\,.
\end{align}
The dominant eigenvalue of this matrix is simpler to compute and yields a particularly simple expression for the current
\begin{align}
    \bar{I}_E^c &\to  \frac{2\Omega\Gamma_c^{\rm eff} \Gamma_h^{\rm eff} (n_c-n_h)}{\Gamma_c^{\rm eff}(1+3 n_c) + \Gamma_h^{\rm eff}(1+3 n_h)}\,.
\label{EQ:AnalIc}
\end{align}
One can see that to achieve cooling of the cold reservoir one needs to have $n_c>n_h$, such that cooling works when $\beta_c\in(\beta_h,3\beta_h)$, which bounds the range of applicability of our engine.
Furthermore, if $\Gamma_{c/h}^{\rm eff}$ scale quadratically (our limit implies that $\Gamma_w^{\rm eff}$ is not the bottleneck either) with the number of two-level systems, so does the current.
For the noise computation we can proceed similarly and obtain
\begin{align}
    \bar{S}_E^c &\to \frac{4\Omega^2 \Gamma_c^{\rm eff} \Gamma_h^{\rm eff}\left[A+ 2 \Gamma_c^{\rm eff} \Gamma_h^{\rm eff} B\right]}{\left[\Gamma_c^{\rm eff} (1+3 n_c) + \Gamma_h^{\rm eff} (1+3 n_h)\right]^3}\,,\nn
    A &\equiv\left[
    \left(\Gamma_c^{\rm eff}(1+3 n_c)\right)^2+   \left(\Gamma_h^{\rm eff}(1+3 n_h)\right)^2 \right]\times\nn
    &\qquad\times(n_c+n_h+2n_c n_h)\,,\nn
    B &\equiv n_c(1+n_c)+n_h(1+n_h)+12 n_c n_h\nn
    &\qquad+ 15 n_c n_h(n_c+n_h)+18 n_c^2 n_h^2\,.
    \label{EQ:AnalSc}
\end{align}

Finally, we remark that similar results arise in the limit where the transitions of the work reservoir are driven by a laser instead, although the proper definition of heat and work for periodically driven master equations is debated~\cite{kalaee2021a}.
In the simplest realization of a resonant laser, the work reservoir is described by the rate matrix
\begin{align}
R_w \to R_L &= \Gamma_L^{\rm eff} \left(\begin{array}{ccc}
    0 & 0 & 0\\
    0 & -1 & +1\\
    0 & +1 & -1
\end{array}\right)\,,
\end{align}
which merely reflects the fact that absorption and emission are equally likely and where $\Gamma_L^{\rm eff}$ is proportional to the pump intensity.
Considering strong driving with $\Gamma_L^{\rm eff}\to\infty$ leads to the same effects as an infinite-temperature work reservoir for current and noise.

\bibliography{Ref_Cooler}

\end{document}